\documentclass[11pt]{article}
\usepackage[utf8]{inputenc}
\usepackage[T1]{fontenc}
\usepackage{geometry}
\usepackage{amsmath}
\usepackage{amsfonts}
\usepackage[english]{babel}
\usepackage[round]{natbib}
\usepackage{bbm}
\usepackage{graphicx}
\usepackage{subcaption}
\usepackage{float}
\usepackage{url}
\usepackage{multicol}
\usepackage[round]{natbib}
\usepackage{color}
\usepackage[dvipsnames]{xcolor}

\definecolor{lava}{rgb}{0.81, 0.06, 0.13}

\usepackage{accents}

\title{Adjusting for both sequential testing and systematic error in safety surveillance using observational data:\\ Empirical calibration and MaxSPRT
}
\author{Martijn J.~Schuemie$^{1,2}$, Fan Bu$^{2}$, Akihiko Nishimura$^{3}$ and Marc A.~Suchard$^{2,4,5}$}
\date{}

\begin{document}

\maketitle

$^{1}$ Observational Health Data Analytics, Janssen Research \& Development, Titusville, NJ \\
$^{2}$ Department of Biostatistics, University of California, Los Angeles, Los Angeles, CA \\
$^{3}$ Department of Biostatistics, Johns Hopkins University, Baltimore, MD \\
$^{4}$ Department of Human Genetics, University of California, Los Angeles, Los Angeles, CA \\
$^{5}$ VA Informatics and Computing Infrastructure, US Department of Veterans Affairs, Salt Lake City, UT \\

\begin{abstract}
Post-approval safety surveillance of medical products using observational healthcare data can help identify safety issues beyond those found in pre-approval trials. 
When testing sequentially as data accrue, maximum sequential probability ratio testing (MaxSPRT) is a common approach to maintaining nominal type 1 error.
However, the true type 1 error may still deviate from the specified one because of systematic error due to the observational nature of the analysis.
This systematic error may persist even after controlling for known confounders.
Here we propose to address this issue by combing MaxSPRT with empirical calibration. 
In empirical calibration, we assume uncertainty about the systematic error in our analysis, the source of uncertainty commonly overlooked in practice.
We infer a probability distribution of systematic error by relying on a large set of negative controls: exposure-outcome where no causal effect is believed to exist.
Integrating this distribution into our test statistics has previously been shown to restore type 1 error to nominal.
Here we show how we can calibrate the critical value central to MaxSPRT.
We evaluate this novel approach using simulations and real electronic health records, using H1N1 vaccinations during the 2009-2010 season as an example. 
Results show that combining empirical calibration with MaxSPRT restores nominal type 1 error. 
In our real-world example, adjusting for systematic error using empirical calibration has a larger impact than, and hence is just as essential as, adjusting for sequential testing using MaxSPRT.
We recommend performing both, using the method described here.
\end{abstract}

\section{Introduction}
When new medical products are brought to market, it is important to continue to monitor their safety.
The phase 2 and 3 clinical trials preceding marketing may uncover most common adverse events attributable to use of these products, but rare adverse events may go undetected due to limited sample size of the trials.
Other adverse events may go undetected if they only affect a subpopulation that was excluded from the clinical trials.
To catch these types of adverse events requires post-marketing drug and vaccine safety surveillance, for example using routinely collected healthcare data, such as administrative claims and electronic health records.
To detect adverse events as quickly as possible, data should be analyzed as they accrue, for example once every month analyzing all data up to that point in time.
It is important to realize that such sequential testing for a safety signal is a form of multiple testing that needs to be adjusted for if one aims to maintain the prespecified type 1 error (i.e.~if one aims to keep the probability of rejecting the null hypothesis of no association between the adverse event and the product exposure when the null is true at the specified $\alpha$-level).
A common approach to adjusting for sequential testing is maximum sequential probability ratio testing (MaxSPRT), often used by the US Food and Drug Administration (FDA) in its safety surveillance \citep{RN171}.

There is, however, another important reason why the true type 1 error rate might deviate from the prespecified $\alpha$, and that is systematic error due to the observational nature of the study \citep{maclure2001causation}.
Systematic error can manifest from multiple sources, including confounding, selection bias, and measurement error.
For example, in a study comparing vaccinated to unvaccinated patients, the vaccinated group may differ from the unvaccinated one in terms of age and fragility that impact the baseline probability of having the adverse event outcome, and failing to adjust for these differences will lead to biased effect size estimates measuring the association.
While there is widespread awareness of the potential for systematic error in observational studies and a large body of research that examines how to diagnose and statistically adjust for specific sources of bias, there is no guarantee that even the best-designed observational study does not contain residual systematic error.

In our prior research we have argued for learning about the potential magnitude and uncertainty of systematic error in an observational study through negative controls \citep{RN10}.
Negative controls are exposure-outcome pairs where no causal effect is believed to exist, and where therefore the true relative risk is assumed to be $1$.
By applying the same study design to negative controls, we can evaluate how far the estimated effect sizes deviate from the truth (i.e., no effect).
Ideally, a negative control also has identical confounding to the exposure-outcome pair of interest \citep{RN179}, but we believe that the true confounding structure is unknowable.
Instead, we propose to use a large sample of negative controls comparable to the exposure-outcome of interest, for example by sharing the same exposure.
Although such a sample does not allow quantifying the exact systematic error in a study, it can be used to infer a distribution of systematic error.
If we assume the systematic error for our exposure-outcome of interest draws from this distribution, we can incorporate it in our statistics to produce calibrated $p$-values \citep{RN10} and confidence intervals \citep{RN6} that demonstrate close to nominal operating characteristics.
Because we use a data-driven approach to estimate the systematic error distribution, we refer to this process as empirical calibration.
Alternatively, one could calibrate against an expert-elicited systematic error distribution.

In this paper we propose to combine adjusting for sequential testing through MaxSPRT with adjusting for residual systematic error using empirical calibration. We first introduce an example using real-world data that will be used throughout the paper.
We subsequently explore the nature of systematic error, and detail how a systematic error probability distribution can be inferred using a set of negative controls.
We demonstrate the impact of accounting for systematic error when performing only a single test (i.e., a non-sequential test), before describing how empirical calibration can be applied to sequential testing. Using simulated and real data, we compare the performance of using both MaxSPRT and empirical calibration to either adjustment by itself, or no adjustment at all.

\section{Running example}
To illustrate systematic error in a real-world setting, we use data from the Evaluating Use of Methods for Adverse Events Under Surveillance (EUMAEUS) project \citep{schuemie2021vaccine}.
Our data source is the Optum\textsuperscript{\texttrademark} de-identified Electronic Health Record dataset (Optum EHR), containing clinical information, prescriptions, lab results, vital signs, body measurements, diagnoses and procedures derived from clinical notes from both inpatient and outpatient environments using natural language processing.
We identify all vaccinations against 2009 pandemic influenza A (H1N1pdm) virus from September 1, 2009 to May 31, 2010.

We further identify a list of negative control outcomes, outcomes not believed to be caused by H1N1pdm vaccines, and, therefore, ideally would not be flagged as a signal by a safety surveillance system.
To identify negative control outcomes that match the severity and prevalence of suspected vaccine adverse effects, a candidate list of negative controls was generated based on similarity of prevalence and percent of diagnoses that were recorded in an inpatient setting (as a proxy for severity).
Manual review of this list by clinical experts created the final list of 93 negative control outcomes (see Supplementary Materials).
Positive controls are outcomes known to be caused by vaccines, and ideally would be detected as signals by a safety surveillance system as early as possible.
However, positive controls are problematic for various reasons \citep{RN51}.
First, vaccine adverse effects that are well established are rare.
Second, even when an effect is established, the magnitude is never known with precision.
Third, for well-established adverse effects, actions are often taken to mitigate the risk, such as careful monitoring or even discontinuation of the vaccine, masking these effects in real-world data.
For these reasons we did not include positive controls in our analyses.

Using two different epidemiological designs we compute incidence rate ratios for each vaccine-negative control outcome pair.
Effect sizes are estimated both at the end of the period, as well as cumulatively at 1-month intervals.

The first design is an historical (background rate) comparator design \citep{black2009importance,klein2021surveillance}, comparing outcomes observed in the 28 days following vaccination to incidence rates of the outcome during the period from September 1, 2008 to May 31, 2009, in the 28 days following one random outpatient visit per person in the database.
We standardize the incidence rates by age and sex to the vaccinated population before computing the incidence rate ratio.

The second design is a self-controlled case series (SCCS) \citep{farrington1995relative,RN180}, comparing the rate of the outcome in the 28 days following vaccination to the rate in the same individuals during all other time in the September 1, 2009 to May 31, 2010 period. 
To avoid bias due to the healthy-vacccinee effect, the 30 days prior to vaccination are excluded from the analysis. Age and season are adjusted for using 5-knot bicubic splines.

\section{Adjusting for systematic error}

\subsection{On the nature of systematic error}

\begin{figure}
     \centering
     \begin{subfigure}[b]{0.45\textwidth}
         \centering
         \includegraphics[width=\textwidth]{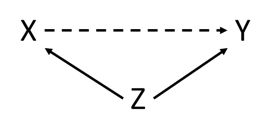}
         \caption{}
         \label{fig: confounding-diagram}
     \end{subfigure}
     \hfill
     \begin{subfigure}[b]{0.5\textwidth}
         \centering
         \includegraphics[width=\textwidth]{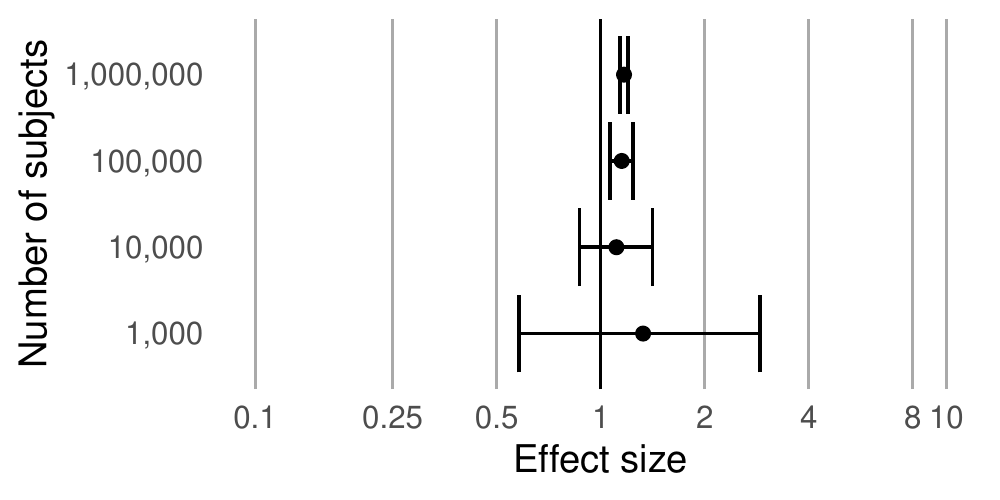}
         \caption{}
         \label{fig: effect-estimates}
     \end{subfigure}
        \caption{Simulating confounding as systematic error. (a) An example directed acyclic diagram showing exposure $X$, outcome $Y$, and confounder $Z$. (b) Effect size estimates when simulating according to the diagram, with no true causal effect and weak confounding, under various sample sizes. }
        \label{fig:sim-confouding}
\end{figure}

As a straight-forward illustration of systematic error, we simulate data under the simple model depicted in Figure~\ref{fig: confounding-diagram}, and defined as
\begin{align*}
    Z &\sim N(0,1),\\
    P(X) &= 0.3 + 0.1 \times Z \text{ and}\\
    P(Y) &= 0.03 + 0.01 \times Z,
\end{align*}
where $Z$ is a continuous confounder, $N(a,b)$ indicates a normal distribution with mean $a$ and variance $b$, $X$ is a binary exposure, and $Y$ is the binary outcome. Probabilities $P(X)$ and $P(Y)$ are truncated to the [0,1] range.
We simulate no causal effect between X and Y, and compute relative risk estimates and 95\% confidence intervals (CI) using logistic regression, for various sample sizes.

Figure~\ref{fig: effect-estimates} shows that, as sample size increases, the CI narrows, and the estimate converges on a value greater than the true simulated causal effect of $1$. We refer to the difference between this asymptotic value and the true effect size as systematic error. In this case it is caused solely by the simulated (weak) confounder, but other systematic causes such as selection bias and measurement error can exist.
In real life, for any exposure-outcome pair, the true systematic error is unknown.
We could assume our study is unbiased, placing 100\% probability that the systematic error is zero, as is the current implied default.
But we argue a more reasonable approach would be to account for our uncertainty about the systematic error.

\subsection{Estimating a systematic error distribution from negative control estimates}

\label{sec: estimate-systematic-error-dist}
Defining a formal probability distribution to account for uncertainty about the bias induced by systematic error in a study is non-trivial.
If we want to reason what the correct distribution must be based on expert knowledge alone, we must understand the sources of bias for which our study design did not sufficiently adjust.
In our experience, people are not very good at guessing how bad the bias in an observational study can be, and tend to severely underestimate bias, especially in a study they themselves designed. An expert-driven process would also lack reproducibility, with different experts likely preferring different systematic error probability distributions.

We therefore use a fully data-driven approach to quantifying uncertainty of the systematic error \citep{RN10, RN74}.
This approach relies on a large set of negative control outcomes, typically between 50 and 100, where we believe the exposure does not cause or prevent the outcome, based on a lack of any evidence in literature, spontaneous reports and product labels \citep{RN71}, and expert review.
We have no requirements on the systematic error that could be associated with each negative control, which we believe to be unknowable. In our running example of H1N1pdm vaccines we identify 93 negative control outcomes, including, e.g., ``contusion of toe'', and ``viral hepatitis C''.

\begin{figure}
    \centering
    \includegraphics[width=\textwidth]{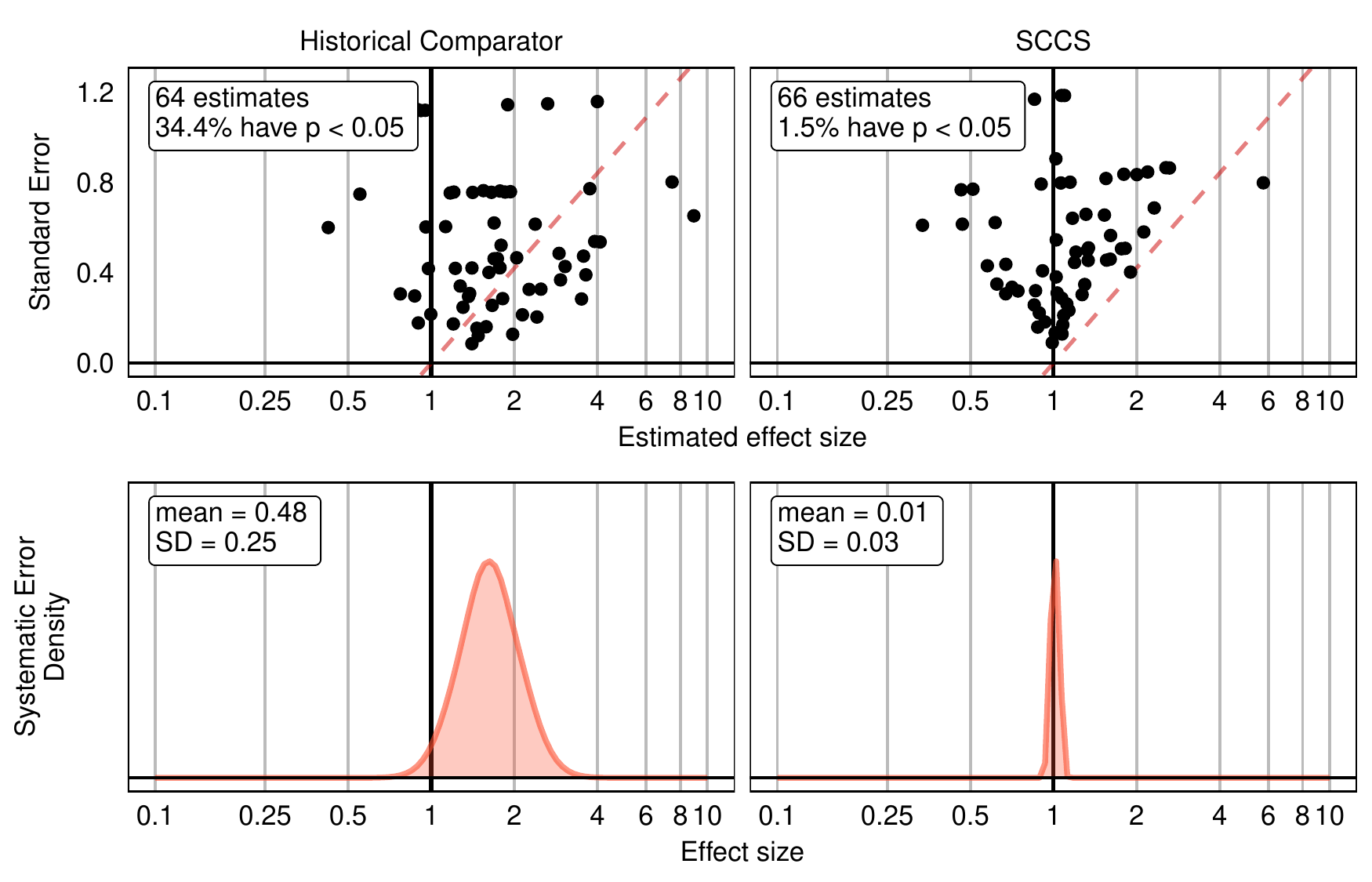}
    \caption{Estimation of the systematic error probability distributions under an historical comparator and self-controlled case series (SCCS) design using negative control estimates.
    In the top plots, each dot represents the effect size estimate for a negative control outcome.
    The red dashed line indicates where the one-sided $p$-value is equal to $0.05$ under a simple $t$-test. Estimates below this line are considered statistically significant (at $\alpha  = 0.05$).
    The bottom plots indicate the systematic error distributions fitted using the negative control estimates.
}
    \label{fig: estimating-error-dist}
\end{figure}

Even though we do not know the systematic error of each individual negative control, we can estimate the distribution of systematic error as illustrated in Figure~\ref{fig: estimating-error-dist}.
We apply our study designs on our data to produce effect size estimates based on maximizing each designs' likelihood function and standard errors (linearly related to the width of the asymptotic CI about the point estimate) for each negative control, or more generally, we can compute the likelihood function for each outcome as a function of effect size.
Note that in this example we were not able to compute estimates for all 93 negative controls, primarily because the outcome is not always observed at least once during the time at risk, leading to an uninformative likelihood function.

As a starting point, we assume the systematic error distribution is Gaussian, and estimate the parameters of this distribution using the negative control estimates.
Formally, for negative control $i$ ($i = 1,\ldots,n$), the likelihood of the parameter of interest $\beta_i$ (e.g. the log incidence rate ratio) is determined by some known  likelihood function $L_i(\beta_i; d_i)$, given the data $d_i$.
Let $\theta_i$ denote the log true effect size (assumed $0$ for negative controls), and let $\tau_i$ denote the systematic error. 
We assume that the parameter of interest $\beta_i$ linearly deviates from the true effect size $\theta_i$, where the addtive deviation term is $\tau_i$, i.e., 
\begin{equation}
	\label{eq: linear-bias-assumption}
	\beta_i = \theta_i + \tau_i. 
\end{equation}
We further assume that the $\tau_i$’s independently arise from a normal distribution with mean $\mu$ and variance $\sigma^2$, i.e., 
$\tau_i \sim N(\mu, \sigma^2)$. 
Given the data $d_i$'s, we can write out the joint likelihood function with respect to $\beta_i$'s, $\tau_i$'s, $\mu$ and $\sigma$:
\begin{equation}
	\label{eq: systematic-error-joint-lik}
	\mathcal{L}(\mu, \sigma, \boldsymbol{\tau}, \boldsymbol{\beta} ; \mathbf{d}) \propto \prod_{i = 1}^n  \left[ L_i(\beta_i; d_i) \varphi( \tau_i \mid \mu, \sigma)\right],
\end{equation}
where $\boldsymbol{\tau} = \{\tau_i\}_{i=1}^n$, $\boldsymbol{\beta} = \{\beta_i\}_{i=1}^n$, $\mathbf{d} = \{d_i\}_{i=1}^n$, and $\varphi$ denotes the normal density function. 

Since for negative control outcomes we believe $\theta_i \equiv 0$ and thus $\beta_i = \tau_i$, by assuming independence among all the analyzed negative controls, we can integrate out the parameters $\beta_i$'s and $\tau_i$'s and thus obtain a marginal likelihood function with respect to only $\mu$ and $\sigma$:
\begin{equation}
	\label{eq: systematic-error-marginal-lik}
	\mathcal{L}(\mu, \sigma; \mathbf{d}) \propto \prod_{i = 1}^n  \int L_i(\tau_i; d_i) \varphi( \tau_i \mid \mu, \sigma)d\tau_i. 
\end{equation}

Maximizing the function above gives us maximum likelihood estimates (MLEs) for $\mu$ and $\sigma$, denoted by 
$\hat\mu$ and $\hat\sigma$. 
Note that this is also a generalization of the approach taken in \cite{RN10}, where the likelihood function $L_i(\beta_i; d_i)$ was restricted to a normal density function. 

An unbiased study design will have a mean $\hat\mu$ and standard deviation $\hat\sigma$ of the estimated systematic error distribution both equal zero (in the limit), meaning that the spread of negative control estimates can be fully explained by the random error expressed in the per-negative-control likelihood functions (e.g. as expressed in their confidence intervals) alone.
As shown in Figure~\ref{fig: estimating-error-dist}, the SCCS design is close to this ideal, but the historical comparator design tends to produce wildly varying levels of systematic error, with an overall tendency to positive systematic error.

We can use the fitted distribution to express our uncertainty about the systematic error in the next exposure-outcome pair for which we wish to estimate the effect size, assuming the new systematic error will draw from the fitted distribution.
Admittedly, the implied assumption of exchangeability is a leap of faith, but we argue that it is less of a leap than asserting with 100\% certainty that no systematic error exists, which is the current status quo.
We can have more confidence in this assumption if we keep either the exposure or the outcome constant across all exposure-outcome pairs (including the negative controls).
Furthermore, prior evaluations of our approach using a leave-one-out design demonstrate that at least within sets of negative controls this exchangeability assumption appears to hold \citep{RN10}.

\subsection{Empirical $p$-value calibration}
\label{sec: p-val-calibration}
By integrating the systematic error probability distribution in our reported statistics about the effect size for the outcome of interest, such as $p$-values and CIs, we can take into account both systematic and random errors simultaneously.

For example, suppose we are interested in a previously unseen effect $\beta_{n+1}$ for an exposure-outcome pair of interest indexed by $n+1$, with true effect size $\theta_{n+1}$ and systematic error induced bias $\tau_{n+1}$. 
Assuming the systematic error term $\tau_{n+1}$ is exchangeable with the biases of previously analyzed $n$ negative controls, we wish to 
perform the following one-sided hypothesis test:
\begin{equation*}
    H_0: \theta_{n+1} = 0, \quad \text{ v.s. } \quad H_1: \theta_{n+1} > 0,
\end{equation*}
which, since $\beta_{n+1} = \theta_{n+1} + \tau_{n+1}$, is equivalent to
\begin{equation}
\label{eq: hypothese-with-tau-v2}
    H_0: \beta_{n+1} = \tau_{n+1}, \quad \text{ v.s. } \quad H_1: \beta_{n+1} > \tau_{n+1}.
\end{equation}

Given the estimated empirical distribution $N(\hat\mu, \hat\sigma^2)$ for $\tau_i$'s, under the null, the estimand $\beta_{n+1}$ shall follow the same normal distribution as $\tau_{n+1}$:
\begin{equation}
    \label{eq: null-distribution}
    \beta_{n+1} \sim N(\hat\mu, \hat\sigma^2) \text{ under } H_0.
\end{equation}

For simplicity, we assume that the likelihood $L(\beta_{n+1}, d_{n+1})$ induces a normal distribution for the MLE $\hat\beta_{n+1}$ with standard deviation $s_{n+1}$ as in \cite{RN10}:
\begin{equation}
    \label{eq: MLE-normality}
    \hat\beta_{n+1} \sim N(\beta_{n+1}, s_{n+1}^2).
\end{equation}

Combining \eqref{eq: null-distribution} and \eqref{eq: MLE-normality}, the marginal distribution for the MLE $\hat\beta_{n+1}$ under the null after integrating out the uncertainty about bias $\tau_{n+1}$ is then
\begin{equation}
    \label{eq: MLE-normality-null}
    \hat\beta_{n+1} \sim N(\hat\mu, \hat\sigma^2 + s_{n+1}^2) \text{ under } H_0.
\end{equation}

Therefore, we can compute a one-sided $p$-value as
\begin{equation*}
   p^{(c)}(\hat\beta_{n+1}) = \Phi\left(\frac{\hat\beta_{n+1} - \hat\mu}{\sqrt{\hat\sigma^2 + s_{n+1}^2}}\right),
\end{equation*}
where $\Phi(\cdot)$ denotes the cumulative distribution function (CDF) of the standard normal.

Since this $p$-value accounts for the uncertainty of bias due to systematic error as well as random sampling error, we refer to it as the ``calibrated $p$-value'' and call such procedure ``empirical calibration.''

\begin{figure}
    \centering
    \includegraphics[width=\textwidth]{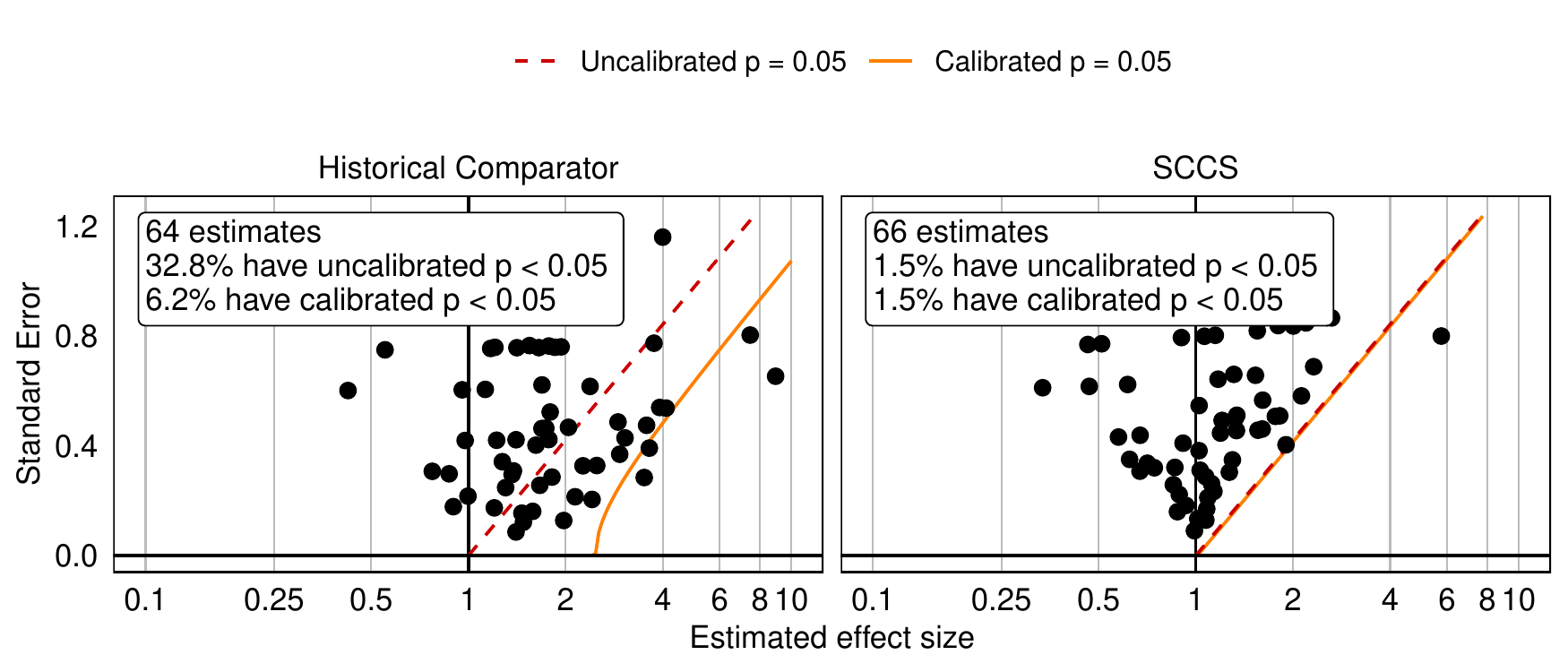}
    \caption{$p$-value calibration under an historical comparator and SCCS design. Each dot represents the effect size estimate for a negative control outcome.
    The red dashed line indicates where the one-sided $p$-value is equal to $0.05$. When ignoring systematic errors, estimates below this line are considered statistically significant at $\alpha  = 0.05$. The orange solid line indicates calibrated one-sided $p^{(c)} = 0.05$.
}
    \label{fig: p-val-calibration}
\end{figure}

Figure~\ref{fig: p-val-calibration} demonstrates $p$-value calibration using the fitted systematic error distributions shown in Figure~\ref{fig: estimating-error-dist}. After calibration, the type 1 error rate is closer to the nominal 5\% for the historical comparator design. For the SCCS design, calibration made little difference, as the type 1 error rate before calibration was already close to nominal.

\section{Adjusting for sequential testing using MaxSPRT}

Drug and vaccine safety surveillance often requires sequential testing of hypotheses against data accruing over time.
This form of multiple testing would lead to inflated type 1 error if unadjusted.
For this reason, MaxSPRT aims to maintain type 1 error across a predefined number of looks at the data.

For a parameter of interest $\beta$ (e.g., a log incidence rate ratio for a specific exposure-outcome pair), MaxSPRT tests the following hypotheses sequentially \citep{RN171}:
\begin{equation}
\label{eq: hypothese-MaxSPRT}
    H_0: \beta = 0, \quad \text{ v.s. } \quad H_1: \beta > 0.
\end{equation}

Under MaxSPRT, at every look of the data at time $t$, we first identify the MLE $\hat\beta_t$ for $\beta$ using the likelihood $L(\beta, d_t)$, and then compute the log likelihood ratio (LLR) defined as follows:
\begin{equation}
    \label{eq: MaxSPRT-LLR}
    LLR_t =
    \begin{cases}
    0 & \hat\beta_t \leq 0\\
    \log\left(\frac{L(\hat\beta_t,d_t)}{L(0,d_t)}\right) & \hat\beta_t > 0 ,
    \end{cases}
\end{equation}
where $d_t$ and $\hat\beta_t$ are the data and MLE at time $t$, respectively.
Note that the LLR is set to $0$ when $\hat\beta_t \leq 0$ because this is a one-sided hypothesis test.

Before initiating the surveillance, a critical value $cv$ is computed based on the type of model, the desired $\alpha$-level, the number of looks that will be made, and the expected sample size at each look.
At every look the $LLR_t$ is compared with the pre-computed $cv$; as soon as $LLR_t > cv$, we reject $H_0$ and declare a signal.

\subsection{Computing the critical value}
\label{sec: Monte-Carlo-cv-computation}
The computation of $cv$ using exact statistics is described in \cite{RN171} and can be performed using the \emph{Sequential} package in \texttt{R}. 
Here we apply a Monte-Carlo approach instead, which we believe is easier to interpret, implement and extend to other models, including models using empirical calibration. 
The original MaxSPRT implements several models when computing the critical values. 
For the historical comparator example method one would typically use the Poisson model implementation, which assumes that under the null the observed outcome counts arise from a background rate that is known with certainty. 
For the SCCS method the recommended model is the binomial model, where under the null the ratio of exposed to unexposed cases is assumed to arise from a known probability (of being exposed). 
Although models that more closely fit our historical comparator and SCCS designs could be implemented, we decided to not change too many things about the MaxSPRT as it is currently used. 
We therefore used the MaxSPRT Poisson model and binomial model for the historical comparator and SCCS designs, respectively.

\subsubsection{Poisson model}
We illustrate the specific computation of $cv$ under the Poisson model here first.
Similar to the original MaxSPRT, we assume the number of looks $T$ and the expected adverse event counts $e_t$ under the null at look $t$ are known. 
Note that $e_t$ is the incremental count, i.e., the additional expected counts since the previous look.
We run $S$ Monte Carlo simulations, for example $S = 10^6$, and for each simulation $s = 1,...,S$ we conduct the following steps:
\begin{enumerate}
	\item At each look $t$, sample the observed event count $o_t^{(s)}$ given the expected count  $e_t$, assuming the null is true:
	\begin{equation}
		o_{t}^{(s)} \sim \text{Poisson}(e_t),
	\end{equation}
	where $\text{Poisson}(\lambda)$ denotes the Poisson distribution with rate $\lambda$;
	\item At each look $t$, compute the log-likelihood ratio $LLR_{t}^{(s)} $ following \eqref{eq: MaxSPRT-LLR}: 
	\begin{equation}
		LLR_{t}^{(s)}  = \log\left(\frac{\text{dPoisson}(\sum_{x=1}^{t} o_{x}^{(s)}; \sum_{x=1}^{t} o_{x}^{(s)})}{\text{dPoisson}(\sum_{x=1}^{x} o_{x}^{(s)}; \sum_{x=1}^{s} e_{x})}\right),
	\end{equation}
	where $\text{dPoisson}(\cdot; \lambda)$ denotes the Poisson probability mass function for a $\text{Poisson}(\lambda)$  distribution;
	\item Take the maximum across all look $t$'s for each simulation:
	\begin{equation}
		LLR_{\max}^{(s)}  = \max_{t=1,...,T} LLR_{t}^{(s)}. 
	\end{equation}
\end{enumerate}

Finally, we select the lowest $cv$ where the fraction of all $LLR_{\max}^{(s)}$'s that exceed the $cv$ is equal to or smaller than the desired significance level $\alpha$.

\subsubsection{Binomial model}
Similar to the original MaxSPRT, we assume that we know the number of looks $T$, the total expected adverse event counts $e_t$ across both exposed and unexposed subjects under the null at look $t$, and the proportion $p$ of exposed subjects among those with adverse events (or in the case of the SCCS, the proportion of patient time considered exposed).
Again, $e_t$ is the incremental count, i.e., the additional expected number of adverse events since the previous look.
Similarly to the Poisson case, we run $S$ Monte Carlo simulations, and for each simulation $s = 1,...,S$ we conduct the following steps:
\begin{enumerate}
	\item At each look $t$, sample the observed number of events  for exposed subjects, $o_t^{(s)}$, given the expected total event count $e_t$ and the proportion of exposed $p$, assuming the null is true:
	\begin{equation}
		o_{t}^{(s)} \sim \text{Binomial}(e_t, p),
	\end{equation}
where $\text{Binomial}(N, q)$ denotes a binomial distribution with total trial size $N$ and incidence probability $q$;
	\item At each look $t$, compute the log-likelihood ratio $LLR_{t}^{(s)} $ following \eqref{eq: MaxSPRT-LLR}: 
	\begin{equation}
		LLR_{t}^{(s)}  = \log\left(\frac{\text{dBinomial}\left(\sum_{u=1}^{t} o_{u}^{(s)}; \sum_{u=1}^{t} e_{u}, \sum_{u=1}^{t} o_{u}^{(s)} / \sum_{u=1}^{t} e_{u}\right)}{\text{dBinomial}\left(\sum_{u=1}^{t} o_{u}^{(s)}; \sum_{u=1}^{s} e_{u}, p\right)}\right),
	\end{equation}
	where $\text{dBinomial}(\cdot; N, p)$ denotes the probability mass function for a  $\text{Binomial}(N, q)$ distribution. 
	\item Take the maximum across all $t$'s for each simulation:
	\begin{equation}
		LLR_{\max}^{(s)}  = \max_{t=1,...,T} LLR_{t}^{(s)}.
	\end{equation}
\end{enumerate}

Finally, we select the lowest $cv$ where the fraction of all $LLR_{\max}^{(s)}$'s that exceed the $cv$ is equal to or smaller than the desired significance level $\alpha$.

\section{Adjusting for both systematic error and sequential testing}
\label{sec:adjusting_for_both}

The hypothesis test MaxSPRT performs can be considered as a special case of the hypothesis test specified in Section~\ref{sec: p-val-calibration} ---
if the distribution for bias term $\tau_{n+1}$ is a normal $N(\mu, \sigma^2)$ distribution with $\mu = \sigma = 0$, then the hypothesis test in \eqref{eq: hypothese-with-tau-v2} is reduced to the form of \eqref{eq: hypothese-MaxSPRT}.
This implies that the original MaxSPRT operates under the assumption that we are 100\% certain that the systematic error induces \emph{exactly zero} bias; in other words, the bias term $\tau$ takes value $0$ with probability $1$, and simply put, the systematic error is completely ignored. 
Now, instead, we allow the possibility of the systematic error inducing nonzero bias, which is captured by the empirical systematic error distribution. Therefore, when the null hypothesis is true, we assume that the estimable log effect $\beta$ equals the bias term $\tau$ that follows a probabilistic distribution which does not necessarily put all its mass at $0$. 
In other words, under $H_0$, the log effect estimand $\beta$ does not simply take on a point value $0$ but rather follows a \emph{null distribution}, defined as the systematic error distribution for the bias term $\tau$.
That is, we can combine empirical calibration with MaxSPRT and perform the following hypothesis test:
\begin{equation}
    \label{eq: hypothesis-test-combined}
     H_0: \beta = \tau, \quad \text{ v.s. } \quad H_1: \beta > \tau,
\end{equation}
where the bias $\tau$ follows a systematic error distribution, which, for simplicity, is assumed to be normal $N(\mu, \sigma^2)$ with mean $\mu$ and variance $\sigma^2$. 

We adjust our procedure for computing the critical value $cv$ by first specifying the systematic error distribution. 
For this, we estimate $\hat{\mu}_t$ and $\hat{\sigma}_t^2$ following the procedure described in Section~\ref{sec: estimate-systematic-error-dist}, by analyzing the negative control estimates obtained at look $t$.
We then sample the systematic error in each simulation $s$ at each look $t$:
\begin{equation}
  \tau_t^{(s)} \sim N(\hat{\mu}_t, \hat{\sigma}_t^2),
\end{equation}
We then sample the data under the null, incorporating the sampled systematic error $\tau_t^{(s)}$ to reflect the adjusted null: 

For the Poisson model, under $H_0: \beta = \tau$, we can set the log rate ratio as $\beta = \tau_t^{(s)}$, and thus the updated sampling step is
\begin{equation}
o_{t}^{(s)} \sim \text{Poisson}(e_t \times \exp(\tau_t^{(s)})).
\end{equation}

For the binomial model, under $H_0: \beta = \tau$, we can set the log odds ratio to $\beta = \tau_t^{(s)}$, which means now the incidence probability $\tilde{p}$ satisfies $\tilde{p}/(1-\tilde{p}) = \exp(\tau_t^{(s)}) \times p/(1-p)$, which means  $\tilde{p} = p \exp(\tau_t^{(s)})/(1+p( \exp(\tau_t^{(s)})-1))$, so the updated sampling step is:
\begin{equation}
o_{t}^{(s)} \sim \text{Binomial}(e_t, \tilde{p}).
\end{equation}

Then we  proceed as before to compute the new, calibrated, critical value $cv_t$.
We note that the critical value $cv_t$ is now a dynamic threshold that changes across looks, as we accrue more data and learn more about the systematic error distribution. 

\section{Simulations}

\subsection{Simulation design}

We examine the operating characteristics of calibrated MaxSPRT first through a synthetic experiment.
For each simulation, we simulate $200$ outcomes, with true effect sizes of $1$, $1.5$, $2$, and $4$ (50 controls per effect size).
A simulation uses one of three systematic error distributions (assumed normally distributed):
\begin{itemize}
    \item mean $\mu = 0$, $\sigma= 0$ (zero expected bias, no uncertainty);
    \item mean $\mu = 0$, $\sigma= 0.2$ (zero expected bias, some uncertainty);
    \item mean $\mu = 0.2$, $\sigma= 0.2$ (positive expected bias, some uncertainty);
\end{itemize}

Each simulation uses either a large or small sample size:
\begin{itemize}
    \item Small: $100,000$ exposed subjects for the historical comparator design, $100$ exposed cases for SCCS arising uniformly in time;
    \item Large: $1,000,000$ exposed subjects for the historical comparator design, $1,000$ exposed cases for SCCS arising uniformly in time.
\end{itemize}

We take $10$ equally-spaced sequential looks in time at the available cases.
Critical values for the MaxSPRT are computed using the actual sample sizes at all looks.
When not using MaxSPRT, a signal is declared when a $p$-value at any look is below the $\alpha$ threshold.
When using empirical calibration, the systematic error distribution is estimated using the simulated negative controls (i.e. those outcomes having true effect size $= 1$), using the data up to the point in time the estimates are computed, thus mirroring what would be done in reality.
Type 1 and 2 error rates are computed with or without MaxSPRT, and with or without empirical calibration, using $\alpha = 0.05$. Each scenario (choice of systematic error distribution, method, and sample size parameter) is repeated $100$ times to produce distributions for type 1 and 2 error rates.
We make our simulation \texttt{R} code available in the Supplementary Material.

\subsection{Simulation results}

\begin{figure}
    \centering
    \includegraphics[width=\textwidth]{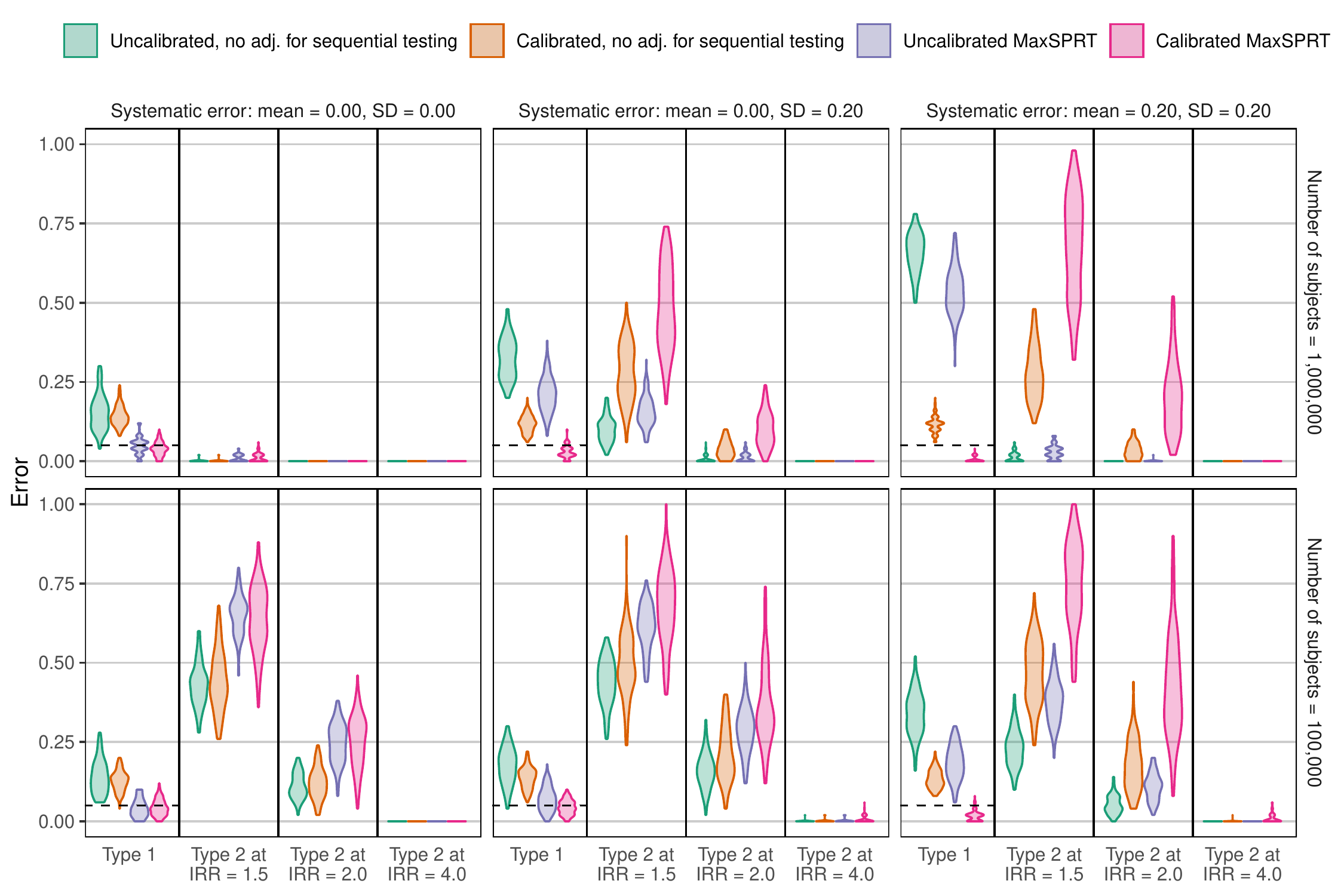}
    \caption{Violin plots showing type 1 and 2 error rates in simulations of the historical comparator design.
    Each panel corresponds to a specific simulation scenario, using a specific sample size (right) and systematic error distribution (top).
    The violin plots show the distribution of type 1 or 2 error rates across the $100$ simulations per simulation scenario.
    Colors indicate what adjustments were used. The dashed line indicates the nominal type 1 error rate (at $\alpha = 0.05$).
}
    \label{fig:sim-hist-comparator}
\end{figure}

\begin{figure}
    \centering
    \includegraphics[width=\textwidth]{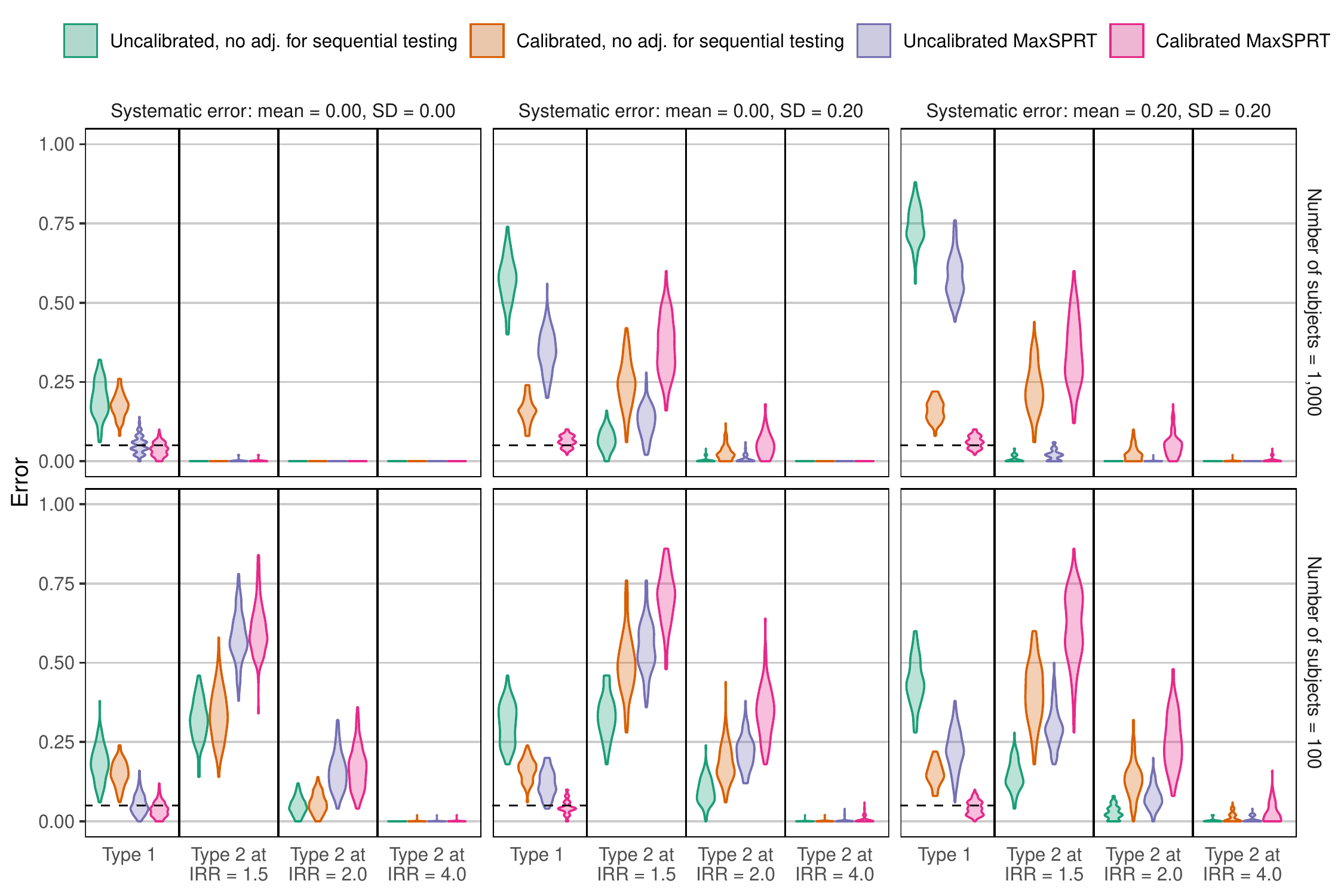}
    \caption{Violin plots showing type 1 and 2 error rates in simulations of the SCCS design.
    Each panel corresponds to a specific simulation scenario, using a specific sample size (right) and systematic error distribution (top).
    The violin plots show the distribution of type 1 or 2 error rates  across the $100$ simulations per simulation scenario.
    Colors indicate what adjustments were used. The dashed line indicates the nominal type 1 error rate (at $\alpha = 0.05$).
}
    \label{fig:sim-SCCS}
\end{figure}

Figures~\ref{fig:sim-hist-comparator} and \ref{fig:sim-SCCS} show the results of our synthetic experiment using the historical comparator design and SCCS design, respectively.
Using MaxSPRT always decreases type 1 error rates, while often increasing type 2 error rates.
When there is no systematic error (mean $\mu = 0$, SD $\sigma= 0$ ), the type 1 and 2 error rates are very similar with or without empirical calibration.
When systematic error is simulated to be present, empirical calibration has a substantial impact, reducing type 1 error while often increasing type 2 error rates.
In these scenarios, the adjustment from empirical calibration is typically much larger than the adjustment for sequential testing.
In all simulation scenarios, the combination of MaxSPRT and empirical calibration achieves a type 1 error rate close to nominal.
In the historical comparator simulations, the mean number of outcomes during the time-at-risk at the end of the study period (after the 10 looks) is 23.1 in the $100,000$ sample size simulations, and 231.0 in the $1,000,000$ sample size simulations.
In the SCCS simulations, the mean number of outcomes during the time-at-risk at the end of the study period is 18.1 in the $100$ sample size simulations, and 181.0 in the $1,000$ sample size simulations.

\section{Real world example}
\subsection{Real-world design}
In our running example of H1N1pdm vaccinations we divide the study period into $9$ calendar months, using the data up to and including a month to compute $p$-values and LLRs.
Critical values for the MaxSPRT are computed using the actual sample sizes at all looks.
When not using MaxSPRT, a signal is declared when a $p$-value at any look is below the alpha threshold. When using empirical calibration, the systematic error distribution is estimated using the negative controls using leave-one-out: for each negative control, calibration uses the systematic error distribution fitted using all other negative controls.
Based on the negative controls, type 1 error rate is computed with or without MaxSPRT, and with or without empirical calibration, using $\alpha = 0.05$.
Since we have only negative control outcomes, we cannot compute type 2 error rates.
The mean number of outcomes during the time at risk at the end of the study period (after 9 months) is 8.5 and 11.7, for the historical comparator and SCCS analyses, respectively.

\subsection{Real-world results}

Table~\ref{tab:real-world-results} shows the type 1 error rates when using the various adjustments.
For the historical comparator design, the type 1 error rate is much larger than nominal when not adjusting for sequential testing and not using empirical calibration.
Here, empirical calibration again has a larger effect than MaxSPRT in moving the type 1 error rate closer to nominal.
For the SCCS design, neither form of adjustment has a large impact on type 1 error.
When combining MaxSPRT with empirical calibration, type 1 error is close to nominal for both designs.

\begin{table}
    \centering
    \caption{Type 1 error rates observed for negative control outcomes of the H1N1pdm vaccine with and without empirical calibration and sequential testing adjustment via MaxSPRT.  Nominal type 1 error rates should approach 5\%.    
    }
    \begin{tabular}{lrr}
    \hline
     & \multicolumn{2}{c}{Type 1 error rate}\\
     & Historical comparator & SCCS \\
     \hline
     Uncalibrated, no adjustment for sequential testing & 28.0\% & 4.3\% \\
     Uncalibrated, MaxSPRT & 18.3\% & 2.2\% \\
     Calibrated, no adjustment for sequential testing & 10.8\% & 5.4\% \\
     Calibrated, MaxSPRT & 5.4\% & 4.3\%\\
     \hline
    \end{tabular}
    \label{tab:real-world-results}
\end{table}

\section{Discussion}

A common concern when performing drug or vaccine safety surveillance is inflated type 1 error due to sequential testing.
However, when using observational data another important concern is increased type 1 error due to systematic error from residual confounding, selection bias, and measurement error.
From our recent experience, systematic error may overshadow inflated type 1 error from sequential testing across several observational designs \citep{schuemie2021vaccine}.
In prior research, we proposed empirical calibration as a reproducible way to account for our uncertainty about the systematic error in an observational study design.
In this paper, we demonstrate how empirical calibration can be combined with adjustment for sequential testing using MaxSPRT.
Our simulation and real-world results show that adjusting for systematic error using empirical calibration often has a stronger impact on maintaining nominal type 1 error than adjusting for sequential testing, and that combining the two types of adjustments leads to close to nominal type 1 error in all evaluated scenarios.

Our simulations show that, when systematic error is simulated to be present, empirical calibration tends to increase type 2 error rates in order to maintain type 1 error rates.
This is similar to adjustment for sequential testing, which also tends to increase type 2 error rates, as there exists an intrinsic trade-off between type 1 and type 2 errors.
Decision makers should take this behavior into account.

In our real-world example, empirical calibration has a large effect for the historical comparator design, but hardly any effect for the SCCS design. 
This results because we observed hardly any systematic error under the SCCS design.
Our simulations suggest that in such a scenario, when type 1 error needs no adjustment, the type 2 error rate is also not increased by the calibration, so although calibration in such a scenario does not help, it also does not hurt.
We point out that, without negative controls and their estimates, we would not have known the specific SCCS design we used in this scenario has little systematic error.
We would also not have guessed that the historical comparator design, despite adjusting for age and sex, and anchoring the historic cohort on outpatient visits to increase comparability, demonstrates substantial systematic error.

Although one could propose a systematic error probability distribution based on expert knowledge of the exposure, outcome, design, and data alone, we advise against it.
Predicting how large systematic error can be after any adjustments, such as matching or using a self-controlled design, is not possible.
In our experience, actual error is more complex than one’s assumptions.
Instead, we propose empirically learning an approximation of the uncertainty by fitting a distribution to our negative control estimates.

Our negative controls are required to have no causal relationship between exposure and outcome, but we make no requirement on the type and magnitude of systematic error because we believe these to be unknowable.
A limitation of our approach is therefore that our negative controls could have very different systematic error than another exposure-outcome pair for which we estimate an effect.
We reduce this concern by using a large sample of negative controls and by having our negative controls share either the exposure or the outcome with the hypothesis of interest.
Although a concern of exchangeability still remains, we believe it is outweighed by the consequence of doing nothing when systematic error is likely present in any observational study.

In conclusion, we propose a novel approach for simultaneously adjusting for sequential testing and performing empirical calibration. Our open-source \emph{EmpiricalCalbration} \texttt{R} package implements this approach, and is available on CRAN. We believe that when doing drug or vaccine safety surveillance, it is important to include negative controls and to perform both types of adjustment.

\section*{Funding}

US Food \& Drug Administration CBER BEST Initiative (75F40120D00039)

\section*{Declaration of Competing Interests}

MJS is an employee of Janssen Research \& Development and shareholder in Johnson \& Johnson.
MAS receives contracts and grants from the US Food \& Drug Administration, the US Department of Veterans Affairs, the US National Institutes of Health and Janssen Research \& Development, the latter two unrelated to the scope of this work.

\newpage
\bibliographystyle{chicago}
\bibliography{ref}

\end{document}